# On the pressure gain of stable flow systems with variable cross-section area

Haocheng Wen and Bing Wang*

School of Aerospace Engineering, Tsinghua University, Beijing, China, 100084

*Corresponding author: wbing@tsinghua.edu.cn

**Abstract:** For the clarification of total pressure gain performance of rotating detonation propulsion systems, the extended Hugoniot curve is proposed and discussed for the stable flow systems with variable cross-section area (SFSVA). The dimensionless pressure integral ($\theta$) along the system walls is found to have critical impact on the pressure gain of the SFSVA with given inlet Mach number ($M_0$) and heat release ($q$). The key to obtain positive pressure gain of a SFSVA is to achieve the matching $\theta$, $M_0$ and $q$.

**Nomenclature**

| | | |
|---|---|---|
| $A$ | = | cross-section area |
| $h$ | = | sensible enthalpy |
| $M$ | = | Mach number |
| $p$ | = | pressure |
| $q$ | = | heat release |
| $s$ | = | entropy |
| $T$ | = | temperature |
| $u$ | = | velocity |
| $\beta$ | = | area ratio of outlet and inlet |
| $\theta$ | = | dimensionless pressure integral |
| $\rho$ | = | density |

Subscripts

| | | |
|---|---|---|
| 0 | = | inlet parameter |
| 1 | = | outlet parameter |

## 1 Introduction

Especially in recent years, detonative propulsions have been a research hotspot in the aerospace field. Three types of detonation wave are adopted in such propulsion systems, namely the pulse detonation [1], standing detonation [2] and rotating detonation [3]. From the perspective of system characteristics, the pulse detonation is a transient system; the standing detonation is a steady system, and the continuously rotating detonation is a specific system with unsteady but stable reactive flows.

Although detonation itself is a pressure gain combustion, the detonative propulsion system does not necessarily achieve a positive total pressure gain. From the stagnation Hugoniot analysis, Wintenberger et al. [4,5] proved that it is impossible for the standing detonation propulsion system to obtain a positive gain. The positive gain of total pressure would be achieved only if the propagating detonation wave applied, namely pulse detonation [4,6]. However, the conclusions are obtained for one-dimensional systems. It is worth discussing whether the positive gain can be achieved in three-dimensional standing detonation systems. As for the rotating detonation system, the unsteady and non-uniform interior flow leads to difficulties in the theoretical analysis and experimental measurements of its total pressure gain. The question of whether the continuously rotating detonation propulsion system can achieve positive gain has caused some controversy in recent years. On the one hand, some theoretical and numerical studies believe that the positive gain can be realized [7–9]. One the other hand, the predicted gain cannot be obtained in experiments [10–12]. To the best knowledge of present authors, the inconsistence is due to the non-uniform definitions of the boundary and inlet parameters of the studied systems. This motivates the present study on clarification of pressure gain of detonative propulsions systems.

To analyze the total pressure gain performance of the rotating detonation and standing detonation systems, the stable flow system with variable cross-section area (SFSVA) is discussed in this study. The extended Hugoniot curve for the SFSVA is proposed and the main influence factors can be analyzed on total pressure gain. Then, the system parameter ranges that can achieve positive gain are shown. Finally, the realization methods of positive gain of rotating detonation and standing detonation systems are discussed.

## 2 A theoretical model of SFSVA
### 2.1 Governing equations for SFSVA

A three-dimensional inviscid SFSVA as shown in Figure 1 is considered in the study, whose inlet and outlet can respectively correspond to the intake and nozzle exit of a real propulsion system. The inflow and outflow parameters of the system are denoted by subscripts 0 and 1, respectively. In the text, $p$, $\rho$, $T$, $M$, $u$ denote static pressure, density, temperature, and Mach number, velocity, respectively. The cross-section areas of the inlet and outlet are $A_0$ and $A_1$. The system internal profile can be monotonically convergent, divergent, or convergent-divergent type, as shown in Figure 1, or the other complex type.

Some assumptions are made to simplify the analysis of SFSVA, as follows:
(1) The system is thermodynamically isolated, but there are internal transfer of technical work and reaction heat release (without the addition of mass), and the net heat release is $q$.
(2) The interior flow of the system can be steady or unsteady, but the inflow and outflow is required to be steady and uniform. Additionally, the inlet and outlet sections are planar and parallel, and the streamlines are parallel to the section normal.
(3) Without loss of generality, the specific heat capacity ratio $\gamma$ is assumed to be constant and equal to 1.4 inside the system for the convenience of entropy calculation, that is $\gamma_0 = \gamma_1 = \gamma = 1.4$.

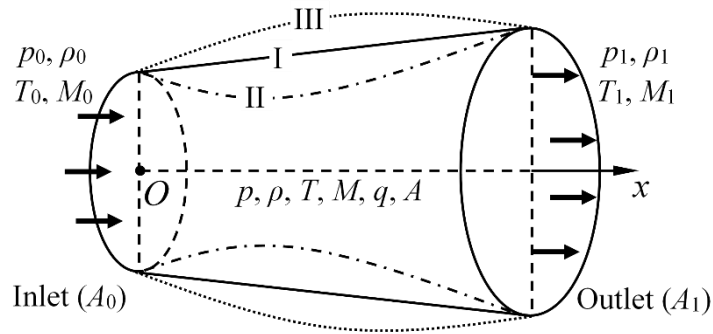

**Figure 1** Schematic of a three-dimensional isolated and inviscid SFSVA.

The conservation governing equations for the inlet and outlet of the SFSVA are as follows,

$$\rho_0 u_0 A_0 = \rho_1 u_1 A_1 \tag{1}$$

$$(p_0 + \rho_0 u_0^2) A_0 + \oiint_\Sigma p n dA = (p_1 + \rho_1 u_1^2) A_1 \tag{2}$$

$$h_0 + \frac{u_0^2}{2} + q = h_1 + \frac{u_1^2}{2} \tag{3}$$

where $h$ is sensible enthalpy and it can be written as $h = \frac{\gamma}{\gamma-1}\frac{p}{\rho}$ for a perfect gas, $\Sigma$ denotes the set of interior walls, $n$ denotes the streamwise component of the wall normal vector.

Define the dimensionless pressure integral $\theta$, coefficient $\alpha$ and area ratio $\beta$ as follows

$$\theta = \frac{\oiint_\Sigma p n dA}{p_0 A_0}, \quad \alpha = \theta + 1, \quad \beta = A_1/A_0$$

Then the momentum equation can be rewritten as

$$\alpha p_0 + \rho_0 u_0^2 = \beta(p_1 + \rho_1 u_1^2) \tag{4}$$

For the one-dimensional system without the variation of cross-section area, the above coefficients satisfy the relations $\beta = 1$, $\theta = 0$, $\alpha = 1$. For the SFSVA, however, $\beta = 1$ may not correspond to $\theta = 0$ due the convergent or divergent walls where the pressure integral can be introduced.

Define new flow parameters as follows,

$$\rho_\alpha \triangleq \rho_0, p_\alpha \triangleq \alpha p_0, u_\alpha \triangleq u_0, \gamma_\alpha \triangleq \gamma_0 = \gamma$$

$$\rho_\beta \triangleq \beta \rho_1, p_\beta \triangleq \beta p_1, u_\beta \triangleq u_1, \gamma_\beta \triangleq \gamma_1 = \gamma$$

It can be proved that

$$h_\alpha = \alpha h_1, \quad h_\beta = h_1, \quad M_\alpha = \frac{1}{\sqrt{\alpha}} M_0, \quad M_\beta = M_1$$

Thus, the governing equations for SFSVA can be rewritten as

$$\rho_\alpha u_\alpha = \rho_\beta u_\beta \tag{5}$$

$$p_\alpha + \rho_\alpha u_\alpha^2 = p_\beta + \rho_\beta u_\beta^2 \tag{6}$$

$$\frac{1}{\alpha} h_\alpha + \frac{u_\alpha^2}{2} + q = h_\beta + \frac{u_\beta^2}{2} \tag{7}$$

The solution of the above equations requires $\theta > -1$ and $\alpha > 0$. When $\theta < -1$, we can define new sets of parameters as follows,

$$\hat{\theta} = -\frac{\oiint_\Sigma p n dA}{p_1 A_1}, \quad \hat{\alpha} = \hat{\theta} + 1, \quad \hat{\beta} = A_0/A_1$$

And the corresponding momentum equation is

$$\hat{\beta}\left(p_0 + \rho_0 u_0^2\right) = \hat{\alpha} p_1 + \rho_1 u_1^2 \tag{8}$$

which is similar to Equation (4). In this study, only the flow process with $\theta > -1$ is discussed. The flow equations with $\theta < -1$ can be solved by the similar method.

Equations (5-7) are similar to the governing equations for the one-dimensional combustion wave [13], and thus the extended Rayleigh line and Hugoniot curve for the SFSVA can be obtained in the similar way.

Define

$$x = \frac{v_\beta}{v_\alpha} = \frac{1}{\beta}\frac{v_1}{v_0}, \quad y = \frac{p_\beta}{p_\alpha} = \frac{\beta}{\alpha}\frac{p_1}{p_0}$$

where $v = 1/\rho$ is the specific volume. The extended Rayleigh line is expressed as

$$y = \left(1 + \gamma_\alpha M_\alpha^2\right) - \gamma_\alpha M_\alpha^2 x \tag{9}$$

The extended Hugoniot curve is expressed as

$$y = \frac{\dfrac{\left(\dfrac{2}{\alpha} - 1\right)\gamma_\alpha + 1}{\gamma_\alpha - 1} - x + 2\tilde{q}}{\dfrac{\gamma_\beta + 1}{\gamma_\beta - 1} x - 1} \tag{10}$$

where the dimensionless heat release is

$$\tilde{q} = \frac{q}{p_\alpha v_\alpha} = \frac{q}{\alpha p_0 v_0} = \frac{\bar{q}}{\alpha}, \quad \text{where } \bar{q} = \frac{q}{p_0 v_0}$$

The solutions for any SFSVA can be expressed as the intersection of extended Rayleigh line and Hugoniot curve. While, the physical solution must meet two requirements,

i. The inlet Mach number $M_\alpha$ is real, that is

$$M_\alpha^2 = \frac{1}{\gamma_\alpha}\frac{y-1}{1-x} > 0 \tag{11}$$

ii. The entropy rise is not less than the theoretical minimum value $\Delta s_{\min}$. The entropy rise is calculated by

$$\Delta s = c_v \ln\left[\frac{p_1/p_0}{\left(\rho_1/\rho_0\right)^\gamma}\right] = \frac{R}{\gamma - 1}\ln\left(\alpha \beta^{\gamma-1} x^\gamma y\right) \tag{12}$$

For the system with the heat release of $q$, $\Delta s_{\min}$ corresponds to the entropy rise through the isothermal heat release process. When the exothermic temperature $T$ is infinite, $\Delta s_{\min}$ approaches 0. Therefore, $\Delta s > 0$ should be met for the physical solution.

In the following discussions, $\gamma$ is assumed to be constant to calculate the system entropy rise, that is $\gamma_0 = \gamma_1 = \gamma$. Finally, the total pressure gain $PG$ of SFSVA is defined as

$$PG = \frac{p_1}{p_0} \frac{\left(1+\frac{\gamma_1-1}{2}M_1^2\right)^{\frac{\gamma_1}{\gamma_1-1}}}{\left(1+\frac{\gamma_0-1}{2}M_0^2\right)^{\frac{\gamma_0}{\gamma_0-1}}} - 1 = \frac{\alpha}{\beta} y \left(\frac{1+\frac{x}{y}\frac{\gamma-1}{2\gamma}\frac{y-1}{1-x}}{1+\alpha\frac{\gamma-1}{2\gamma}\frac{y-1}{1-x}}\right)^{\frac{\gamma}{\gamma-1}} - 1 \quad (13)$$

### 2.2 Solutions for SFSVA without heat release ($q = 0$)

Assuming $q = 0$, $\beta = 2$ and $\gamma = 1.4$, the extended Hugoniot curves and the corresponding flow parameters for the SFSVA with $\alpha = 0.7$ and 1.5 are shown in Figure 2. The Hugoniot curves pass above and below the point $(\xi, \eta) = (1, 1)$ respectively for $\alpha = 0.7$ and 1.5. The typical solutions for non-reactive quasi-one-dimensional flow can be found in Figure 2, such as

i. When $\alpha = 0.7$ and $v_1/v_0 = 1.0$, the solved parameters are $p_1/p_0 = 1.44$, $M_0 = 1.82$, $M_1 = 0.74$ and $PG = -0.63$, corresponding to the process in a divergent nozzle with high ambient pressure, where the supersonic flow overexpands and decelerates to subsonic.

ii. When $\alpha = 1.5$ and $v_1/v_0 = 2.4$, the solved parameters are $p_1/p_0 = 0.13$, $M_0 = 1.75$, $M_1 = 2.52$ and $PG = -0.04$, corresponding to the process in a divergent-convergent nozzle with low ambient pressure, where the supersonic flow first decelerate to $M = 1$ and then accelerates to supersonic.

When $\alpha = 1.5$ and $v_1/v_0 > 2.5$, $PG$ is positive but $\Delta s$ is negative, indicating it is impossible to achieve the positive $PG$ in the SFSVA without heat release.

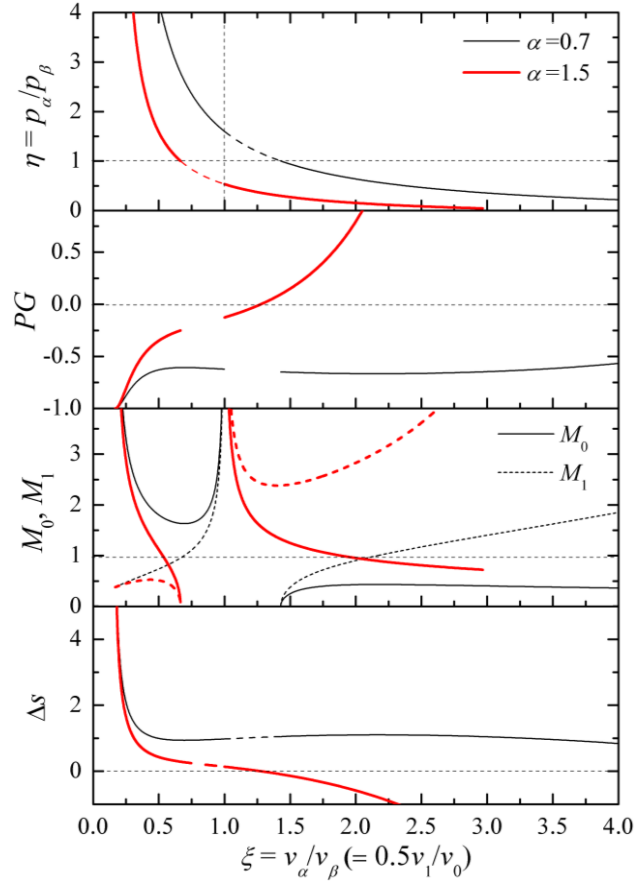

**Figure 2** Hugoniot curves and the corresponding solutions for the SFSVA without heat release ($q = 0$) under different $\alpha$. $\beta = 2$, $\gamma = 1.4$.

## 2.3 Pressure gain performance for SFSVA with heat release ($q > 0$)

For any SFSVA with $\beta \neq 1$, it is always possible to add an isentropic, non-reactive and divergent /convergent flow process before the system inlet or after the outlet, to obtain a new combined system with $\beta = 1$. For the new system, $PG$ remains unchanged, while an additional pressure integral $\Delta \theta$ is introduced by the additional flow process, and thus $\alpha$ is changed. $\Delta \theta$ can be solved by the isentropic relationship and momentum conservation Equation (2). Obviously, the pressure gain performance of the combined system with $\beta = 1$ can represent the performance of any SFSVA. The specific implement approach of the physical solution is not discussed in this section. Only the parametric influence on $PG$ and the parameter range for positive $PG$ are analyzed here.

Given $q = 1$, $\beta = 1$ and $\gamma = 1.4$, the solutions for $\theta = -0.3$, 0.1 and 0.5 (that is $\alpha = 0.7$, 1.1 and 1.5) are shown in Figure 3. When $\alpha$ is small ($\alpha = 0.7$ and 1.1), the Hugoniot curves pass above the point $(\xi, \eta) = (1, 1)$. For the solutions at the upper branch of Hugoniot curve ($\xi < 1$, $M_\alpha > 1$), $PG$ is always negative, and the local maximum $PG$ is obtained when $M_0$ takes the minimum value (where $M_1 = M_\beta = 1$, corresponding to the Chapman-Jouguet detonation solution). For the solutions at the lower branch ($\xi > 1$, $M_\alpha < 1$), $PG$ is minimum when $M_0$ takes the maximum value (where $M_1 = M_\beta = 1$, corresponding to the Chapman-Jouguet deflagration solution). In the region where $\xi > 1$, $M_\alpha < 1$ and $M_\beta > 1$, $PG$ increases as $M_0$ decreases. When $M_0$ increases to a certain critical value, $\Delta s$ is equal to 0 and the maximum $PG_{max}$ for the SFSVA is reached. For $\alpha = 0.7$, the positive $PG$ is only obtained when $\xi > 1$, $M_\alpha < 1$ and $M_\beta > 1$. When $\alpha$ increases to 1.1, $PG$ is always positive at the entire lower branch.

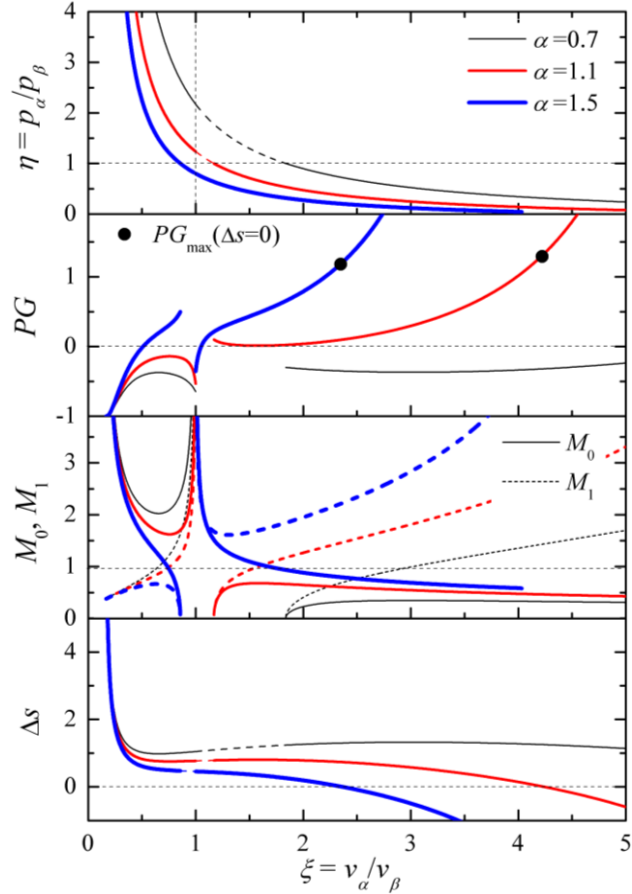

**Figure 3** Hugoniot curves and the corresponding solutions for the SFSVA with heat release ($q > 0$) under different $\alpha$. $\beta = 1$, $\gamma = 1.4$.

When $\alpha$ is large ($\alpha = 1.5$), the Hugoniot curve passes below the point (1, 1). The positive *PG* can be achieved at both the upper and lower branches, and the corresponding $M_0$ can be subsonic or supersonic, and *PG* increases as $M_0$ decreases.

It should be noted that for the one-dimensional flow, the weak detonation solution ($\xi < 1$, $M_\alpha > 1$ and $M_\beta > 1$) and strong deflagration solution ($\xi > 1$, $M_\alpha < 1$ and $M_\beta > 1$) do not exist. However, these two solutions can be realized in the SFSVA. For the strong deflagration solution, if the system has a convergent-divergent nozzle, heat can be added into the subsonic flow in the convergent section, and then the flow can be accelerated to supersonic in the convergent-divergent nozzle. Similarly, the weak detonation solution is also achievable. Additionally, Figure 3 indicates that it is impossible to achieve positive *PG* for the supersonic inflow in the one-dimensional system ($\alpha = 1$ and $\beta = 1$), which is consistent with the conclusion of Wintenberger et al. [5], however, it is possible in the SFSVA when $\alpha$ is large ($\alpha = 1.5$).

Figure 4a further illustrates the influence of the dimensionless heat release $\hat{q}$ on the $M_0$ range that can achieve the positive *PG* in the SFSVA with $\alpha > 1$ (or $\theta > 0$) and $\beta = 1$. The dimensionless heat release is defined by $\hat{q} = \bar{q}/\bar{q}_{cr}$, where

$$\bar{q}_{cr} = \frac{\gamma}{\gamma-1}(\alpha-1) = \frac{\gamma}{\gamma-1}\theta \tag{14}$$

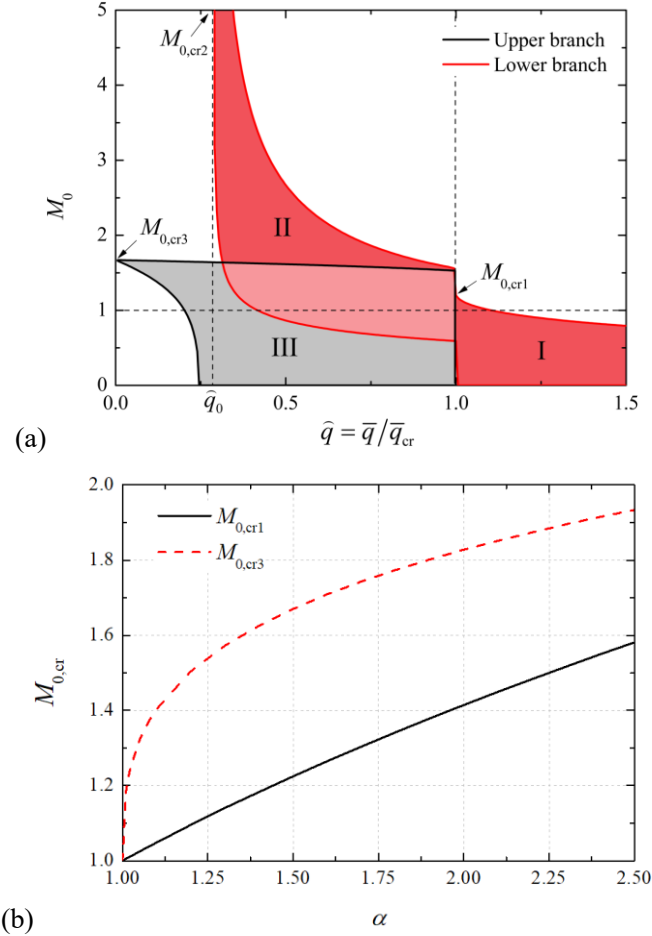

(a)

(b)

**Figure 4** (a) The $M_0$ range for positive *PG* under different $\hat{q}$ with $\alpha = 1.5$; and (b) Variation of $M_{0,cr1}$ and $M_{0,cr3}$ with $\alpha$. $\beta = 1$.

Figure 4a can be divided into three regions. In region I where $\hat{q} > 1$, it can be proved from Equation (10) that $\eta > 1$ when $\xi = 1$, indicating the Hugoniot curve passes above the point (1, 1). As shown in Figure 3, the positive $PG$ can be obtained only when $\eta < 1$ and $M_\alpha < 1$. When $\hat{q} = 1$, it can be proved that $M_\alpha = 1$, and the critical inlet Mach number $M_{0,\text{cr1}} = \sqrt{\alpha}$.

In regions II and III where $\hat{q} < 1$, the Hugoniot curve passes below the point (1, 1) and the positive $PG$ can be achieved at both the upper and lower branches. In the overlap region of II and III, the inlet $M_0$ is equal, while the outlet flow parameters are different. The dimensionless heat release in region II satisfies $\hat{q}_0 < \hat{q} < 1$, where $\hat{q}_0 = \theta/\bar{q}_{\text{cr}} = 1 - 1/\gamma$. When $\hat{q} = \hat{q}_0$, the Hugoniot curve passes the point $(\xi, \eta) = (1, 1/\alpha)$, where the critical Mach number $M_{0,\text{cr2}}$ approaches infinity. In region III, $0 < \hat{q} < 1$. When $\hat{q} = 0$, the critical Mach number $M_{0,\text{cr3}}$ can be solved by combining $PG = 1$ and Equation (10). The variation curves of $M_{0,\text{cr1}}$ and $M_{0,\text{cr3}}$ with $\alpha$ are shown in Figure 4b. Obviously, when $\alpha$ approaches 1, both $M_{0,\text{cr1}}$ and $M_{0,\text{cr3}}$ approach 1.

Figure 4 indicate that with the given heat release $q$, the required $M_0$ for the positive $PG$ is mainly related to $\alpha$ (or $\theta$). When $\alpha > 1$ (or $\theta > 0$), the positive $PG$ can be always achieved if $M_0$ matches with $q$ and $\theta$.

The influence of $\theta$ on $PG$ is illustrated in Figure 6 with $\hat{q} = 1.2$. Given a specific $M_\alpha$, if $M_\beta < 1$, $PG$ increases as $\theta$ increases. Particularly, when $M_\alpha = 0$, $PG = \alpha/\beta - 1$. If $M_\beta > 1$, $PG$ decreases as $\theta$ increases, however, the maximum possible $PG$ is much higher for the greater $\theta$. Therefore, increasing $\theta$ is generally beneficial to improve the system pressure gain performance.

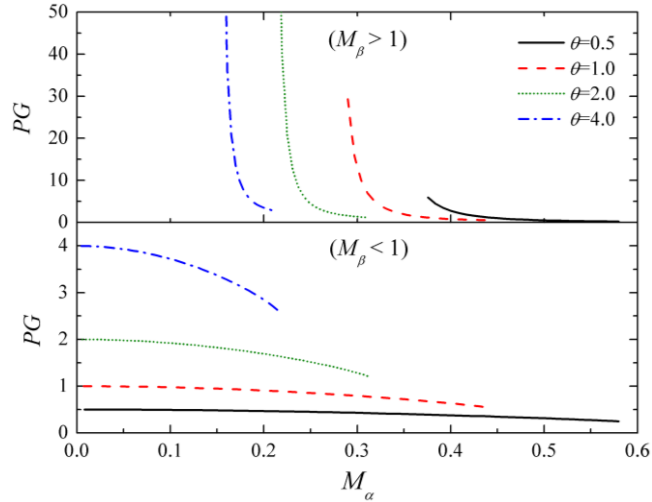

**Figure 5** Effects of $\theta$ and $M_\alpha$ on $PG$ ($\hat{q} = 1.2$). $\beta = 1$.

## 3 Realization Methods of positive *PG* in special SFSVAs

The analysis in Section 2 suggests that the key to obtain a positive $PG$ in SFSVA is to achieving a matching $\theta$ with the given $M_0$ and $q$. The realization methods of positive $PG$ in the rotating detonation system (unsteady system) and quasi-one-dimensional steady system (such as standing detonation system) are discussed in this section.

### 3.1 Rotating detonation system

Generally, the inflow and outflow of a rotating detonation system are unsteady and non-uniform. Before applying the extended Hugoniot curve to analyze the rotating detonation system, the flow filed needs to be equivalent as follows,

1) Assume that the rotating detonation system is axisymmetric with a small channel width, the interior flow is stable and the radial component is uniform. Then the system can be unrolled to a quasi-two-dimensional zone as shown in Figure 6.
2) The cross-section area of the combustor is constant and equal to 0, and thus the interior dimensionless pressure integral is 0.
3) An anti-backflow zone (ABZ) is attached upstream of the combustor, and the disturbance generated by the rotating detonation wave is suppressed and cut off at a certain section in this zone. The inlet of ABZ corresponds to the system inlet, where the flow is steady and uniform, and its area is equal to $A_0$. The pressure integral introduced in ABZ is equal to $\theta_0$.
4) An ideal mixer is attached downstream of the combustor, in which the non-uniform flow completes an isentropic process to become steady and uniform at the outlet of mixer (that is the system outlet). The area of the outlet is also equal to $A_0$. After the isentropic process, the total pressure is unchanged, but an additional pressure integral is introduced and equal to $\theta_1$.

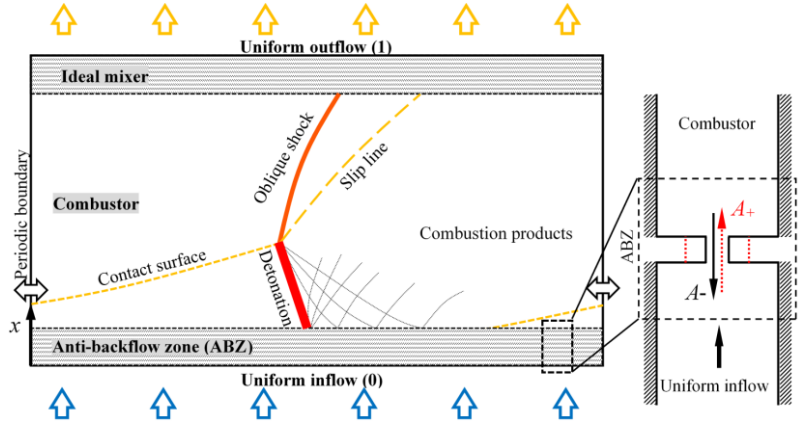

**Figure 6** Schematic of the unrolled quasi-two-dimensional rotating detonation system, and an abstract structure of ABZ with a throat.

After the above equivalence, the extended Hugoniot curve can be applied to analyze the inlet and outlet flow parameters of rotating detonation system. The equivalent system has an area ratio $\beta = 1$ and dimensionless pressure integral $\theta = \theta_0 + \theta_1$.

The value of $\theta_0$ depends on the specific structure of ABZ. An abstract structure of ABZ with a throat is considered as shown in Figure 6. Its inlet and outlet area are equal to $A_0$. The throat area is $A+$ when the upstream and downstream pressure gradient $\Delta p > 0$, otherwise the area is $A-$. Obviously, for different values of $A+$ and $A-$, the value of $\theta_0$ is different. Consider two special cases: (1) when $A+ = A- = A_0$, $\theta_0 = 0$; (2) when $A+ = A_0$ and $A- = 0$, $\theta_0 > 0$. For the other general cases, $\theta_0$ may be greater or less than 0. For examples, in some experiments of rotating detonation engines [12,14], the injector at the combustor head can be equivalently regarded as a kind of ABZ that satisfies $A+ = A- \ll A_0$. In most cases, $\theta_0$ is much less than 0 because the average static pressure at the ABZ inlet (corresponding to the reactant plenum) is much higher than that at the outlet (corresponding to the combustor).

Based on the momentum conservation equation, the expression of $\theta_1$ is as follow

$$\theta_1 = \frac{p_1 + \rho_1 u_1^2}{p_0} - \frac{\oiint_{\Sigma_{IM}} (p + \rho u^2) dA}{p_0 A_0} \tag{15}$$

where $\Sigma_{IM}$ denotes the mixer inlet. The calculation method of $p_1, \rho_1$ and $u_1$ can refer to the Appendix. Equation (15) shows that $\theta_1$ is related to the distribution of pressure, density, and velocity after the rotating detonation wave, which is determined by the outlet parameters of ABZ.

Since both $\theta_0$ and $\theta_1$ are related to ABZ, it is possible to adjust the value of $\theta_0$ and $\theta_1$ by designing a specific ABZ structure to obtain the matching $\theta$ with the given $M_0$ and $q$, and thus achieve the positive $PG$ in the rotating detonation system.

### 3.2 Quasi-one-dimensional steady system

For a quasi-one-dimensional steady system, its flow parameter is only the function of streamwise coordinate $x$. By solving the differential equations for mass, momentum and energy, we can obtain the following equations

$$\frac{d\rho}{dx} = -\frac{\rho}{u}\frac{du}{dx} - \frac{\rho}{A}\frac{dA}{dx} \tag{16}$$

$$\frac{dp}{dx} = -\rho u \frac{du}{dx} \tag{17}$$

$$\frac{du}{dx} = \frac{(\gamma-1)q\frac{d\lambda}{dt} - u\frac{c^2}{A}\frac{dA}{dx}}{c^2(1-M^2)} \tag{18}$$

where $0 \leq \lambda \leq 1$ is the heat release progress variable, $t$ is time, and $c$ is sound speed. The derivation of the above equations is similar to the solving process of equations for ideal ZND model [13]. Based on Equations (16-18), all flow variables can be expressed as the function of $d\lambda/dt$ and $dA/dx$. The required $\theta$ for positive $PG$ can be first determined by the given $M_0$ and $q$, and then it can be realized by designing the matching expressions of $d\lambda/dt$ and $dA/dx$.

For the steady standing detonation propulsion system, the inlet $M_0$ is generally greater than 4 and the expression of $d\lambda/dt$ is definite. Theoretically, the positive $PG$ can be achieved by designing a special wall profile to match with the required $\theta$.

### 4 Conclusions

The total pressure gain performance of the SFSVA is investigated based on the extended Hugoniot curve. The possible approaches to achieve the positive gain of rotating detonation system and standing detonation system are proposed. The main conclusions include:

The total pressure gain of SFSVA is dependent on the inlet $M_0$, heat release $q$ and dimensionless pressure integral $\theta$. In order to achieve the positive gain, $M_0$, $q$ and $\theta$ must meet the specific matching relation. Specially, if $M_0 > 1$, the positive gain exists only when $\theta > 0$.

The analysis on the stable rotating detonation system indicates that the anti-backflow zone has a critical impact on the total pressure gain. It is necessary to obtain the matching $\theta$ by designing the structure of anti-backflow zone to achieve the positive gain at given $M_0$ and $q$.

In order to achieve the positive gain of quasi-one-dimensional steady system (including the standing detonation system), the expressions of the heat release rate $d\lambda/dt$ and the profile varying rate $dA/dx$ have to be designed to obtain the matching $\theta$.

**Appendix: Flow parameters at the outlet of ideal mixer**

Assume that the distributions of flow parameters including $p$, $\rho$, $u$ at the mixer inlet are given. According the isentropic relation, we can obtain

$$\oiint_{\Sigma_{IM}} \rho u \ln\left(\frac{p}{\rho^\gamma}\right) dA - \rho_1 u_1 \ln\left(\frac{p_1}{\rho_1^\gamma}\right) A_0 = 0 \tag{A1}$$

Additionally, the mass and energy conservation equations are as follows,

$$\oiint_{\Sigma_1} \rho u \, dA - \rho_1 u_1 A_0 = 0 \tag{A2}$$

$$\oiint_{\Sigma_{IM}} \left(\frac{\gamma}{\gamma-1}\frac{p}{\rho} + \frac{u^2}{2}\right) dA - \left(\frac{\gamma}{\gamma-1}\frac{p_1}{\rho_1} + \frac{u_1^2}{2}\right) A_0 = 0 \tag{A3}$$

In the above equations, $p_1$, $\rho_1$ and $u_1$ are unknown variables, which can be solved by combining Equations (A1-A3).